# A Lagrangian for the quantionic field equation

Samir Lipovaca
slipovaca@nyse.com

Keywords: quantions**,** Lagrangian, field, group, algebra

*Abstract:* The purpose of this paper is to present a Lagrangian from which we can derive the quantionic field equation written in the Dirac gauge using the principle of stationary action.

**(1) Introduction**

    Structural unification of relativity and quantum mechanics has been intractable since 1920s. Relatively new approach to the structural unification is presented in the form of the algebra of quantions [1].This mathematical structure extends non relativistic quantum mechanics to a relativistic theory by generalizing its underlying number system (the field of complex numbers). The most important results derived as theorems from the algebra of quantions are related to the invariance groups and to the equations of motions (field equations). For example, in standard physics, the Lorentz group (the local Minkowski metric) is the observed structure of spacetime. In the quantionic approach, this metric is an intrinsic property of quantions. In the Standard Model, the properties of elementary particles are encapsulated in the gauge group $SU(3)xSU(2)xU(1)$ where $SU(2)xU(1)$ group defines the electroweak unification. In the quantionic approach, there is an intrinsic gauge group, the quantionic gauge group, denoted by $U_q(1)$. This group is analogous to the group of phase transformations $U(1)$ in the field of complex numbers. Formulated in terms of complex numbers, the group $U_q(1)$ is isomorphic with the group $SU(2)xU(1)$. Four vector potentials appear as differential connections in the formation of the quantionic covariant derivation operator $D - iH$. They are also generators of the quantionic gauge group $U_q(1)$. It seems these potentials have the properties needed to be interpreted as the vector potentials of the electroweak theory. In quantionic physics, Born interpretation of the function $\rho = \psi^*\psi$ as a probability density is naturally generalized and replaced by Zovko's interpretation which leads to Dirac and Schrodinger equations.

    The quantionic field equation written in the Dirac gauge,

$$[D - iH]|q) = im\gamma^1|q^*) \qquad (1)$$

is taken as the starting point of all subsequent investigations. The mass of the quantionic field $|q)$ is $m$, $\gamma^1$ is Dirac's matrix, and $|q^*)$ is the complex conjugate of $|q)$. The four component submatrices $H_\alpha$ in the decomposition

$$H = R^\mu \Lambda_\mu = H_0 + H_1 + H_2 + H_3$$

are differential gauge connections for the four Abelian subgroups of the quantionic gauge group $U_q(1)$. Coefficients $R^0$ to $R^3$ are Hermitian R-type quantions. $\Lambda_\mu$ form a basis of $4\times 4$ Hermitian matrices which is an extension of Pauli's basis of sigma matrices to the quantionic left algebra $\alpha$.

It is unclear if the principle of stationary action is really needed since in quantionic physics Dirac's equation can be obtained from quantions [2]. In a field theory context we may be forced to start from a Lagrangian. From the point of view of quantum field theory, we start with a Lagrangian density and derive equations of motions from the Lagrangian density using the principle of stationary action. In this paper we will present a Lagrangian density from which we can derive the quantionic field equation (1) using the principle of stationary action.

We will start with a Lagrangian when $H = 0$. Then a global symmetry of the Lagrangian is revealed. Next, we make this symmetry local. In an analogy to the Yang-Mills theory we modify the Lagrangian to include the case when $H \neq 0$. Last, we finish with Discussion and Conclusions.

**(2) Lagrangian when $H = 0$**

To simplify the problem, let's assume $H = 0$ which reduces the quantionic field equation to

$$D|q) = iA|q^*) \qquad (2)$$

where $A = m\gamma^1$. It is easy to verify that $A^2 = -m^2$. Multiplying eq (2) on the left by $A$, we obtain

$$AD|q) = -im^2|q^*). \qquad (3)$$

This equation hints a Lagrangian density $\alpha_A$ as

$$\alpha_A = -\frac{1}{m^2}(q|AD|q) + \frac{i}{2}(q^*|q) - \frac{i}{2}(q|q^*). \qquad (4)$$

The quantionic fields $|q)$ and $|q^*)$ are represented by $4\times 1$ complex matrices

$$|q) = \begin{pmatrix} q_1 \\ q_2 \\ q_3 \\ q_4 \end{pmatrix}, |q^*) = \begin{pmatrix} q_1^* \\ q_2^* \\ q_3^* \\ q_4^* \end{pmatrix}$$

and

$$(q| = \begin{pmatrix} q_1^* & q_2^* & q_3^* & q_4^* \end{pmatrix}, (q^*| = \begin{pmatrix} q_1 & q_2 & q_3 & q_4 \end{pmatrix}$$

are Hermitian conjugates of $|q)$ and $|q^*)$. The quantionic derivation operator $D$ is a Hermitian differential matrix operator given by

$$D = \begin{pmatrix} d & 0 & \delta & 0 \\ 0 & d & 0 & \delta \\ \delta^* & 0 & \Delta & 0 \\ 0 & \delta^* & 0 & \Delta \end{pmatrix}$$

and

$$d = \partial_0 + \partial_3, \quad \Delta = \partial_0 - \partial_3, \quad \delta = \partial_1 + i\partial_2, \quad \delta^* = \partial_1 - i\partial_2$$

in terms of partial derivatives. Applying simple rule of matrix multiplication, the Lagrangian density $\alpha_A$, or the Lagrangian for simplicity, is

$$\alpha_A = -\frac{1}{m^2} q_a^* A_{ab} D_{bc} q_c + \frac{i}{2} q_a q_a - \frac{i}{2} q_a^* q_a^*. \tag{5}$$

Any index $a, b, c$, etc. that appears twice in a product is assumed to be summed from 1 to 4. We define the action, $S$, as:

$$S = \int d^4 x \, \alpha_A.$$

Requiring the action to be stationary, that is, $\delta S = 0$ for $|q)$, $|q^*)$ varying in spacetime arbitrarily, generates the field equations from $\alpha_A$. The variations in $|q)$, $|q^*)$ are, however, taken to vanish at the extremities of the spacetime integration. $\delta \alpha_A$ is given by

$$\delta \alpha_A = -\frac{1}{m^2} \delta q_a^* A_{ab} D_{bc} q_c - \frac{1}{m^2} q_a^* A_{ab} \delta D_{bc} q_c + i q_a \delta q_a - i q_a^* \delta q_a^*. \tag{6}$$

Since $\delta D_{bc} = D_{bc} \delta$, the second term in eq (6) becomes

$$-\frac{1}{m^2} q_a^* A_{ab} D_{bc} \delta q_c = -\frac{1}{m^2} D_{bc}(q_a^* A_{ab} \delta q_c) + \frac{1}{m^2} D_{bc}(q_a^* A_{ab}) \delta q_c$$

which yields

$$\delta \alpha_A = \delta q_a^* \left[ -\frac{1}{m^2} A_{ab} D_{bc} q_c - i q_a^* \right] + \left[ \frac{1}{m^2} D_{bc}(q_a^* A_{ab}) + i q_c \right] \delta q_c - \frac{1}{m^2} D_{bc}(q_a^* A_{ab} \delta q_c).$$

Thus

$$\delta S = \int d^4 x \delta q_a^* \left[ -\frac{1}{m^2} A_{ab} D_{bc} q_c - i q_a^* \right] +$$

$$+ \int d^4 x \left[ \frac{1}{m^2} D_{bc}(q_a^* A_{ab}) + i q_c \right] \delta q_c - \qquad (7)$$

$$- \int d^4 x \frac{1}{m^2} D_{bc}(q_a^* A_{ab} \delta q_c) = 0.$$

We can rewrite the last term in eq (7) as

$$\int d^4 x \frac{1}{m^2} D_{bc}(q_a^* A_{ab} \delta q_c) = \frac{1}{m^2} \int d^4 x \partial_\mu j^\mu$$

where the four-current $j^\mu = \frac{1}{2} Tr(Q^+ \delta q \Lambda_\mu)$ and

$$Q = \begin{pmatrix} -q_4 & q_2 & 0 & 0 \\ -q_3 & q_1 & 0 & 0 \\ 0 & 0 & -q_4 & q_2 \\ 0 & 0 & -q_3 & q_1 \end{pmatrix}, \delta q = \begin{pmatrix} \delta q_1 & \delta q_3 & 0 & 0 \\ \delta q_2 & \delta q_4 & 0 & 0 \\ 0 & 0 & \delta q_1 & \delta q_3 \\ 0 & 0 & \delta q_2 & \delta q_4 \end{pmatrix}.$$

In general, Gauss' theorem tell us that $\int_V d^4 x \partial_\mu j^\mu = \int_{\partial V} dS_\mu j^\mu$ where $dS_\mu$ is the outward pointing normal vector to the surface $\partial V$ which bounds $V$. Since we assume that $\delta q = 0$ at the surface $\partial V$ the last term in eq (7) vanishes. Requiring the first two terms in eq (7) to vanish for $\delta q_a$, $\delta q_a^*$ arbitrary functions of spacetime yields the following equations:

$$-\frac{1}{m^2} A_{ab} D_x q_c - i q_a^* = 0$$

$$\frac{1}{m^2} D_x (q_a^* A_{ab}) + i q_c = 0.$$

(8)

In terms of matrices the first equation can be expressed as $-\frac{1}{m^2} AD|q) = i|q^*)$. Multiplying this matrix equation on the left by $A$, using the identity $A^2 = -m^2$, we obtain the quantionic field equation (2). Since $A$ is an antisymmetric matrix, the second equation can be manipulated as:

$$\frac{1}{m^2} (D^T)_{cb} (q_a^* - A_{ba}) + i q_c =$$

$$-\frac{1}{m^2} (D^T)_{cb} (A_{ba} q_a^*) + i q_c =$$

$$-\frac{1}{m^2} D^T A|q^*) + i|q) = 0.$$

(9)

Multiplying the last equation on the left by $A$ we obtain $-\frac{1}{m^2} A D^T A|q^*) = -iA|q)$. It is an easy task to verify $A D^T A = -m^2 D^\#$ so the second equation is:

$$D^\# |q^*) = -iA|q)$$

(10)

where

$$D^\# = \begin{pmatrix} \Delta & 0 & -\delta & 0 \\ 0 & \Delta & 0 & -\delta \\ -\delta^* & 0 & d & 0 \\ 0 & -\delta^* & 0 & d \end{pmatrix}.$$

To summarize, in this section we derived 2 quantionic field equations from the Lagrangian (4) using the principle of stationary action:

$$D|q) = iA|q^*)$$
$$D^\# |q^*) = -iA|q).$$

(11)

Notice the second quantionic field equation does not involve $D^*$ which we would obtain if we complex conjugate eq (2). Surprisingly, if we look at another quantionic field $|Q)$ related to $|q)$ via $A$,

$$|Q) = A|q), |Q^*) = A|q^*),$$

and we replace $|q)$, $|q^*)$ in the second equation of (11) by $|Q)$ and $|Q^*)$ respectively, then we obtain the needed equation $D^*|Q^*) = -iA|Q)$, since $|q^*) = -\frac{1}{m^2}A|Q^*)$, $AD^\#A = -m^2 D^*$ and $A^* = A$. On the other hand,

$$D^*|q^*) = -iA|q),$$
$$D^\#|q^*) = -iA|q)$$

implies

$$D^*|q^*) = D^\#|q^*). \tag{12}$$

In the Appendix A we show that the quantionic field must necessary satisfy the Laplace equation in two dimensions $x^1$ and $x^3$ in order that eq (12) is valid:

$$(\frac{\partial^2}{\partial x^{1^2}} + \frac{\partial^2}{\partial x^{3^2}})|q) = 0.$$

**(3) A global symmetry of $\alpha_A$**

Let us suppose $|q)$, $|q^*)$ transform under some matrix $U$ as follows:

$$|q) \to U|q), |q^*) \to U|q^*).$$

Hence,

$$(q| \to (q|U^+, (q^*| \to (q^*|U^+$$

and the Lagrangian $\alpha_A$ becomes

$$-\frac{1}{m^2}(q|U^+ADU|q) + \frac{i}{2}(q^*|U^+U|q) - \frac{i}{2}(q|U^+U|q^*).$$

Obviously the following conditions must be satisfied if we require that $\alpha_A$ is invariant under $U$:

$$U^+ADU = AD$$
$$U^+U = I. \qquad (13)$$

Thus $U$ must be some unitary $4x4$ matrix. In addition we assume $U$ does not depend on spacetime coordinates, so it is a global unitary transformation. As a further restriction we assume that $D$ acts on $U$ as a linear derivation operator on a constant, leading to $U^+AUD = AD$ and

$$U^+AU = A. \qquad (14)$$

It is easy to show that the following matrix $U$ obeys eq (14):

$$U = \begin{pmatrix} V & 0 \\ 0 & \sigma_1 V \sigma_1 \end{pmatrix} \qquad (15)$$

where $V$ is an unitary $2x2$ matrix. In order that $D(U|q\rangle) = UD|q\rangle$, $U$ must be of the form

$$U = \begin{pmatrix} a & 0 \\ 0 & a \end{pmatrix}$$

meaning that $V$ must commute with $\sigma_1$, since $\sigma_1^2 = I$. Starting with a general expression for an unitary $2x2$ matrix

$$V = \begin{pmatrix} e^{i(\alpha-\frac{\beta}{2}-\frac{\delta}{2})}\cos\frac{\gamma}{2} & -e^{i(\alpha-\frac{\beta}{2}+\frac{\delta}{2})}\sin\frac{\gamma}{2} \\ e^{i(\alpha+\frac{\beta}{2}-\frac{\delta}{2})}\sin\frac{\gamma}{2} & e^{i(\alpha+\frac{\beta}{2}+\frac{\delta}{2})}\cos\frac{\gamma}{2} \end{pmatrix}$$

where $\alpha, \beta, \gamma, \delta$ are some real numbers, it follows that $V$ must be of the form

$$V = e^{i\alpha}\begin{pmatrix} \cos\frac{\gamma}{2} & i\sin\frac{\gamma}{2} \\ i\sin\frac{\gamma}{2} & \cos\frac{\gamma}{2} \end{pmatrix} \qquad (16)$$

in order to commute with $\sigma_1$. Thus a general unitary matrix $U$ which leaves the Langrangian $\alpha_A$ invariant is

$$U = \begin{pmatrix} V & 0 \\ 0 & V \end{pmatrix}$$

where $2x2$ unitary matrix $V$ is given by eq (16). Since $V$ is

$$V = e^{i\alpha} T$$

where $T = \begin{pmatrix} \cos\frac{\gamma}{2} & i\sin\frac{\gamma}{2} \\ i\sin\frac{\gamma}{2} & \cos\frac{\gamma}{2} \end{pmatrix}$ is an arbitrary $2x2$ unimodular, $\det T = 1$, unitary matrix, the general unitary matrix $U$ can be also expressed as

$$U = e^{i\alpha} \begin{pmatrix} T & 0 \\ 0 & T \end{pmatrix}.$$

Hence $U$, which leaves the Langrangian $\alpha_A$ invariant, is an element of the quantionic gauge group $U_q(1)$ which is isomorphic with the group $SU(2) \times U(1)$ since $e^{i\alpha} \in U(1)$ and $T \in SU(2)$.

**(4) Making $U$ symmetry of $\alpha_A$ local**

Let us assume now the matrix $U$ from the previous section is local, meaning $\alpha$ and $\gamma$ of the matrix $V$ depend on spacetime coordinates $x^\mu$. We require the Lagrangian $\alpha_A$ to be invariant under this local $U$. In other words, we want to construct local gauge theory based on $\alpha_A$. Hence,

$$(q|U^+ADU|q) = (q|U^+A(DU)|q) + (q|U^+AUD|q) \tag{17}$$

since the quantionic derivation operator $D$ satisfies the Leibniz identity

$$D(U|q)) = (DU)|q) + UD|q). \tag{18}$$

Using eq (14), eq(17) becomes

$$(q|U^+ADU|q) = (q|U^+A(DU)|q) + (q|AD|q). \tag{19}$$

Obviously the first term on the right side of eq (18) does not exist in $\alpha_A$, so the Lagrangian $\alpha_A$ must be modified in such a way to compensate this term. In an analogy to the Yang-Mills theory, we replace $D|q)$ by the quantionic covariant derivative

$$D^c|q) = (D - \alpha M)|q) \tag{20}$$

where the constant $\alpha$ and matrix $M$ will be determined later. We demand that $D^c|q)$ transforms under $U$ transformations exactly as $|q)$ itself:

$$(D - \alpha M')U|q) = U(D - \alpha M)|q). \tag{21}$$

Then expanding eq (20) using the Leibniz identity (18) and cancelling $U(D|q))$ term on both sides we find

$$[(DU) - \alpha M'U]|q) = -\alpha UM|q). \tag{22}$$

Hence eq (22) must be satisfied for any $|q)$,

$$(DU) - \alpha M'U = -\alpha UM. \tag{23}$$

Finally, multiplying eq (23) on the right by $U^+$ and using the fact that $U$ is an unitary matrix, one must have the following transformation law for the matrix $M$:

$$M' = UMU^+ + \frac{1}{\alpha}(DU)U^+ \tag{24}$$

in order for the following Lagrangian to be invariant under $U$

$$\alpha_A \to L = -\frac{1}{m^2}(q|AD^c|q) + \frac{i}{2}(q^*|q) - \frac{i}{2}(q|q^*). \tag{25}$$

It can be shown that $(q|AD^c|q)$ is indeed invariant under $U$:

$$(q|AD^e|q) = (q|AD|q) - \alpha(q|AM|q) \rightarrow$$
$$\rightarrow (q|U^+ADU|q) - \alpha(q|U^+AM'U|q) =$$
$$= (q|U^+A(DU)|q) + (q|U^+AUD|q) - \alpha(q|U^+A(UMU^+ + \frac{1}{\alpha}(DU)U^+)U|q) =$$
$$= (q|U^+A(DU)|q) + (q|AD|q) - \alpha(q|AM|q) - (q|U^+A(DU)|q) =$$
$$= (q|AD - \alpha AM|q) = (q|AD^e|q).$$

**(5) Lagrangian when $H \neq 0$**

Written in matrix components the Lagrangian (25) is

$$L = -\frac{1}{m^2} q_a^* A_{ab} D_{bc} q_c + \frac{\alpha}{m^2} q_a^* A_{ab} M_{bc} q_c + \frac{i}{2} q_c q_c^* - \frac{i}{2} q_c^* q_c^*. \quad (26)$$

Requiring the action to be stationary, that is $\delta S = 0$, we obtain

$$\delta S = \int d^4 x \delta L = \int d^4 x \delta q_a^* [-\frac{1}{m^2} A_{ab} D_{bc} q_c + \frac{\alpha}{m^2} A_{ab} M_{bc} q_c - i q_a^*] +$$
$$+ \int d^4 x [\frac{1}{m^2} D_{bc}(q_a^* A_{ab}) + \frac{\alpha}{m^2} q_a^* A_{ab} M_{bc} + i q_c] \delta q_c - \quad (27)$$
$$- \int d^4 x \frac{1}{m^2} D_{bc}(q_a^* A_{ab} \delta q_c) = 0.$$

The last term in eq (27) is identical to the last term in eq (7). Therefore it can be written as $\int_{\partial V} dS_\mu j^\mu$. Since we assume that $\delta q = 0$ at the surface $\partial V$, the last term in eq (27) vanishes. Requiring the first two terms in eq (27) to vanish for $\delta q_a$, $\delta q_a^*$ arbitrary functions of spacetime yields the following equations:

$$-\frac{1}{m^2} A_{ab} D_{bc} q_c + \frac{\alpha}{m^2} A_{ab} M_{bc} q_c - i q_a^* = 0$$
$$\frac{1}{m^2} D_{bc}(q_a^* A_{ab}) + \frac{\alpha}{m^2} q_a^* A_{ab} M_{bc} + i q_c = 0. \quad (28)$$

The first equation in (28) can be expressed as

$$-\frac{1}{m^2} A_{ab}(D_{bc} - \alpha M_{bc})q_c = iq_a^*. \tag{29}$$

The term in the ( ) brackets is the quantionic covariant derivative $D^c$ and eq (29) in the matrix form is

$$-\frac{1}{m^2} AD^c|q) = i|q^*). \tag{30}$$

Multiplying eq (30) on the left by $A$ yields

$$D^c|q) = (D - \alpha M)|q) = iA|q^*)$$

or

$$D|q) = \alpha M|q) + iA|q^*). \tag{31}$$

Eq (31) is identical to the quantionic field equation (1) if and only if

$$\alpha = i, \, M = H. \tag{32}$$

Hence our Lagrangian is

$$L = -\frac{1}{m^2}(q|A(D - iH)|q) + \frac{i}{2}(q^*|q) - \frac{i}{2}(q|q^*). \tag{33}$$

Now the second equation of (28) is

$$\frac{1}{m^2} D_{bc}(q_a^* A_{ab}) + \frac{i}{m^2} q_a^* A_{ab} H_{bc} + iq_c = 0. \tag{34}$$

Since $A$ does not depend on spacetime and $D_{bc}$ is a linear differential operator, (34) becomes

$$\frac{A_{ab}}{m^2}(D_{bc} + iH_{bc})q_a^* + iq_c = 0$$

or

$$\frac{A_{ab}}{m^2}(D^{c+})_{bc}q_a^* + iq_c = 0 \tag{35}$$

where $D^{c+} = D^+ + iH^+ = D + iH$ is Hermitian conjugate of the quantionic covariant derivative $D^c$. Using the fact that $A$ is an antisymmetric matrix, (35) can be simplified further as

$$\frac{A_{ab}}{m^2}(D^{c*})_{cb}q_a^* + iq_c = -\frac{1}{m^2}(D^{c*})_{cb}A_{ba}q_a^* + iq_c = 0$$

or in the matrix form

$$-\frac{1}{m^2}D^{c*}A|q^*) = -i|q). \tag{36}$$

Multiplying eq (36) on the left by $A$ yields

$$-\frac{1}{m^2}AD^{c*}A|q^*) = -iA|q). \tag{37}$$

It can be shown that $AD^{c*}A = -m^2(D^\# - iH^\#) = -m^2 D^{c\#}$ and eq (37) becomes

$$(D^\# - iH^\#)|q^*) = -iA|q)$$

or

$$D^\#|q^*) = iH^\#|q^*) - iA|q). \tag{38}$$

This quantionic field equation does not involve $D^*$ which we would obtain if we complex conjugate eq (1). Similarly to the section 2, if we look at another quantionic field $|Q)$ related to $|q)$ via $A$,

$$|Q) = A|q), |Q^*) = A|q^*),$$

and we replace $|q)$, $|q^*)$ in eq (38) by $|Q)$ and $|Q^*)$ respectively, then we obtain the needed equation

$$D^*|Q^*) = -iH^*|Q^*) - iA|Q)$$

since $AD^\# A = -m^2 D^*$ and $AH^\# A = m^2 H^*$. Then,

$$(D^* + iH^*)|q^*) = -iA|q)$$
$$(D^\# - iH^\#)|q^*) = -iA|q)$$

implies

$$(D^* + iH^*)|q^*) = (D^\# - iH^\#)|q^*). \tag{39}$$

In the Appendix B we show that there exists a quantionic field $|Q^0)$ which necessary satisfies the Laplace equation in two dimensions $x^1$ and $x^3$ in order that eq (39) is valid:

$$\left(\frac{\partial^2}{\partial x^{1^2}} + \frac{\partial^2}{\partial x^{3^2}}\right)|Q^0) = 0.$$

**(6) Discussion**

A technique for arriving at the equations of motion in classical mechanics is the use of a Lagrangian and a variational principle. It is possible to derive the field equations using an analogous approach based on a Lagrangian. In quantionic physics there is the quantionic field equation (1) which is taken as the starting point of all subsequent investigations. An interesting question is : What Lagrangian would give the quantionic field equation using the principle of stationary action? We showed the Lagrangian (33) is such a Lagrangian.

If the Lagrangian is invariant under a continuous group of transformations, then there exist locally conserved quantities. These quantities, like the electric charge, flow in space-time and are described in terms of currents. This is in essence Noether's theorem. We showed Lagrangian (33) is invariant under both, global and local, $U_q(1)$ transformations. Since the symmetry is continuous, we can consider an infinitesimal change $\delta q_a$, $\delta q_a^*$. Since the Lagrangian (33) does not change, we have

$$0 = \delta L = \delta q_a^*\left[-\frac{1}{m^2}A_{ab}D_{bc}q_c + \frac{i}{m^2}A_{ab}H_{bc}q_c - iq_a^*\right] +$$
$$+ \left[\frac{1}{m^2}D_{bc}(q_a^*A_{ab}) + \frac{i}{m^2}q_a^*A_{ab}H_{bc} + iq_c\right]\delta q_c - \tag{40}$$
$$- \frac{1}{m^2}D_{bc}(q_a^*A_{ab}\delta q_c)$$

Terms in $[\ ]$ brackets vanish due to the equations of motion and we obtain

$$0 = -\frac{1}{m^2} D_{bc}(q_a^* A_{ab} \delta q_c) = -\frac{1}{m^2} \partial_\mu j^\mu.$$

We showed in the section (2) $j^\mu = \frac{1}{2} Tr(Q^+ \delta q \Lambda_\mu)$. Hence, $\delta L = 0 \Rightarrow \partial_\mu j^\mu = 0$. We have found a conserved current!

It is rather unexpected to demand that $|q^*)$ transforms under $U_q(1)$ transformations like $|q)$ instead of $U^*|q^*)$. The alternative would lead to an orthogonality condition $U^T U = 1$ but the transformation law (14) for $A$ would include $U^+$ instead of $U^T$. An intriguing point is that the Lagrangian $L$ does not lead directly to the complex conjugate of eq (1), but to eq (38) which involves $D^\#$. Now if we look at another quantionic field $|Q)$ related to $|q)$ via $A$,

$$|Q) = A|q), |Q^*) = A|q^*),$$

and we replace $|q), |q^*)$ by $|Q)$ and $|Q^*)$ in eq (38) respectively, then we obtain the complex conjugate of eq (1). Needless to say, this duality relation via $A$ between two equations is also in effect when $H = 0$. Since the equations have the same term $-iA|q)$, eq (39) must be valid (eq (12) when $H = 0$). This implies that there exists a quantionic field $|Q^0)$ which necessary satisfies the Laplace equation in two dimensions $x^1$ and $x^3$. $|Q^0)$ coincides with $|q)$ when $H = 0$. We showed in the Appendix A and B when the full azimuthal range is permitted in $x^1$, $x^3$ plane, then $|q)$, which satisfies eq (1), can be expanded in terms of the solutions of this Laplace equation justifying eq (39) and eq (12). We may be a little disturbed with this azimuthal symmetry, since from the relativistic point of view we are thus singling out $x^1$ and $x^3$ coordinates. Thus we must raise a question is there a Lagrangian which would give the quantionic field equation (1) and only its complex conjugate via the principle of stationary action? A research effort related to this question is in progress.

Going back to the Lagrangian $L$, the first term is a covariant kinetic term. Unlike the covariant kinetic term $i\bar{\psi} D \psi$ of the Standard Model, it contains the matrix $A$ which characterizes the mass of the quantionic field $|q)$. $D - iH$ is a covariant derivative, so that $H$ plays the role of a differential connection or potential. However, the Lagrangian $L$ does not contain a kinetic term for $H$. Thus $L$ represents a gauge-less limit of a full Lagrangian, where potentials of $H$ are external non propagating fields. The other two terms in $L$ are similar to the mass term in the Lagrangian for Dirac's equation $\bar{\psi}(iD - m)\psi$. Thus an open problem is the search for the full Lagrangian that would include the kinetic term for $H$ and other terms similar to the Standard Model Lagrangian, for example the potential term $V$ in the fields $\Phi$, $\Phi^+$. We might expect from this Lagrangian to address spontaneous symmetry breaking in quantionic physics.

**(7) Conclusions**

In this paper we have presented the Lagrangian $L$ from which we derived the quantionic field equation (1) written in the Dirac gauge using the principle of stationary action. An intriguing point is that this Lagrangian does not lead directly to the complex conjugate of eq (1), but to an equation of motion which involves $D^{\#}$. It seems an azimuthal symmetry in $x^1$, $x^3$ plane for the quantionic field $|q\rangle$ is needed. From the relativistic point of view we are thus singling out $x^1$ and $x^3$ coordinates. Thus a question is raised is there a Lagrangian which would give the quantionic field equation (1) and only its complex conjugate via the principle of stationary action? Next, the Lagrangian $L$ does not contain a kinetic term for $H$. $L$ represents a gauge-less limit of a full Lagrangian, where potentials of $H$ are external non propagating fields. Hence an open problem is the search for the full Lagrangian similar to the Standard Model Lagrangian.

**Appendix A**

Since in the $2 \times 2$ block matrix form

$$D^* = \begin{pmatrix} d & \delta^* \\ \delta & \Delta \end{pmatrix}, \quad D^{\#} = \begin{pmatrix} \Delta & -\delta \\ -\delta^* & d \end{pmatrix}$$

and if we assume

$$|q^*\rangle = \begin{pmatrix} Q_1^* \\ Q_2^* \end{pmatrix}$$

in the $2 \times 1$ block matrix form, then eq (12) yields the following equations:

$$(d - \Delta)Q_1^* + (\delta^* + \delta)Q_2^* = 0,$$
$$(\delta + \delta^*)Q_1^* + (\Delta - d)Q_2^* = 0.$$

Replacing $d - \Delta$ by $2\partial_3$ and $\delta + \delta^*$ by $2\partial_1$, leads to

$$\frac{\partial Q_1^*}{\partial x^3} + \frac{\partial Q_2^*}{\partial x^1} = 0,$$

$$\frac{\partial Q_1^*}{\partial x^1} - \frac{\partial Q_2^*}{\partial x^3} = 0. \tag{A1}$$

Applying $\dfrac{\partial}{\partial x^1}$ to the first equation and $\dfrac{\partial}{\partial x^3}$ to the second equation, we obtain the Laplace equation in 2 dimensions for $Q_2^*$:

$$\left(\partial_1^2 + \partial_3^2\right)Q_2^* = 0.$$

Similarly, we obtain the Laplace equation in 2 dimensions for $Q_1^*$ applying $\dfrac{\partial}{\partial x^3}$ to the first equation and $\dfrac{\partial}{\partial x^1}$ to the second equation:

$$\left(\partial_1^2 + \partial_3^2\right)Q_1^* = 0.$$

Thus the quantionic fields $|q^*\rangle$, $|q\rangle$ must obey the Laplace equation:

$$\left(\partial_1^2 + \partial_3^2\right)|q^*\rangle = 0,$$
$$\left(\partial_1^2 + \partial_3^2\right)|q\rangle = 0.$$

Obviously, it is enough to consider only one of the fields, for example $|q\rangle$, since the results for the other field are obtained by complex conjugation. Every component of the $|q\rangle$ obeys the Laplace equation:

$$\left(\partial_1^2 + \partial_3^2\right)q_i = 0, \, i = 1,\ldots,4.$$

or in polar coordinates

$$\frac{1}{\rho}\frac{\partial}{\partial \rho}\left(\rho\frac{\partial q_i}{\partial \rho}\right) + \frac{1}{\rho^2}\frac{\partial^2 q_i}{\partial \phi^2} = 0$$

where

$$\rho = \sqrt{x_1^2 + x_3^2}, \, \phi = \tan^{-1}\left(\frac{x^3}{x^1}\right).$$

The general solution of the Laplace equation in two dimensions [3] is of the form

$$q_i = a_0 + b_0 \ln \rho + \sum_{n=1}^{\infty} a_n \rho^n \sin(n\phi + \alpha_n) + \sum_{n=1}^{\infty} b_n \rho^{-n} \cos(n\phi + \beta_n)$$

when there is no restriction on $\phi$ and in order that $q_i$ is single-valued. Since the origin is included $(\rho = 0)$, all the $b_n$ are zero in order that $q_i$ is finite. Therefore,

$$q_i = a_0(x^0, x^2) + \sum_{n=1}^{\infty} \rho^n A_n(x^0, x^2) \sin(n\phi) + \sum_{n=1}^{\infty} \rho^n B_n(x^0, x^2) \cos(n\phi) \quad (A2)$$

where

$$a_0(x^0, x^2) = \frac{1}{2\pi} \int_0^{2\pi} q_i(x^0, x^2, 0, \phi) d\phi,$$

$$A_n(x^0, x^2) = \frac{1}{\pi \rho_0^n} \int_0^{2\pi} \sin(n\phi) q_i(x^0, x^2, \rho_0, \phi) d\phi,$$

$$B_n(x^0, x^2) = \frac{1}{\pi \rho_0^n} \int_0^{2\pi} \cos(n\phi) q_i(x^0, x^2, \rho_0, \phi) d\phi$$

assuming that values of $q_i$ at $\rho = 0$ and $\rho = \rho_0$ are known. Finally, (A1) is an additional condition on $q_i$ leading to

$$q_1 = a_0(x^0, x^2) + \sum_{n=1}^{\infty} \rho^n A_n(x^0, x^2) \sin(n\phi) + \sum_{n=1}^{\infty} \rho^n B_n(x^0, x^2) \cos(n\phi),$$

$$q_3 = b_0(x^0, x^2) + \sum_{n=1}^{\infty} \rho^n B_n(x^0, x^2) \sin(n\phi) + \sum_{n=1}^{\infty} \rho^n (-A_n(x^0, x^2)) \cos(n\phi).$$

(A3)

Notice up to a sign the exchange $A_n \leftrightarrow B_n$ of the coefficients. Similar expressions are for $q_2$ and $q_4$.

**Appendix B**

Let again

$$|q^*) = \begin{pmatrix} Q_1^* \\ Q_2^* \end{pmatrix}$$

in the $2 \times 1$ block matrix form, so eq (39) yields the following equations:

$$2\frac{\partial Q_1^*}{\partial x^3} + 2\frac{\partial Q_2^*}{\partial x^1} = -i(r+s)\sigma_3 Q_1^* - i(c^*-c)\sigma_3 Q_2^*$$

$$2\frac{\partial Q_1^*}{\partial x^1} - 2\frac{\partial Q_2^*}{\partial x^3} = -i(c-c^*)\sigma_3 Q_1^* - i(r+s)\sigma_3 Q_2^*.$$

(B1)

Obviously there exist such functions $F^*$ and $G^*$ which are solutions of the following partial differential equations of the first order:

$$\frac{\partial F^*}{\partial x^3} = \frac{i}{2}(r+s)\sigma_3 Q_1^*, \quad \frac{\partial F^*}{\partial x^1} = \frac{i}{2}(c-c^*)\sigma_3 Q_1^*$$

$$\frac{\partial G^*}{\partial x^3} = -\frac{i}{2}(r+s)\sigma_3 Q_2^*, \quad \frac{\partial G^*}{\partial x^1} = \frac{i}{2}(c^*-c)\sigma_3 Q_2^*.$$

Hence, (B1) becomes

$$\frac{\partial Q_1^{0*}}{\partial x^3} + \frac{\partial Q_2^{0*}}{\partial x^1} = 0$$

$$\frac{\partial Q_1^{0*}}{\partial x^1} - \frac{\partial Q_2^{0*}}{\partial x^3} = 0$$

(B2)

where $Q_1^{0*} = Q_1^* + F^*$ and $Q_2^{0*} = Q_2^* + G^*$. The quantionic fields $|Q^{0*})$, $|Q^0)$ must obey the Laplace equation:

$$(\partial_1^2 + \partial_3^2)|Q^{0*}) = 0,$$
$$(\partial_1^2 + \partial_3^2)|Q^0) = 0.$$

The treatment for these fields is the same as in the Appendix A. Hence, in analogy to (A3)

$$q_1^0 = a_0(x^0,x^2) + \sum_{n=1}^{\infty} \rho^n A_n(x^0,x^2)\sin(n\phi) + \sum_{n=1}^{\infty} \rho^n B_n(x^0,x^2)\cos(n\phi),$$

$$q_3^0 = b_0(x^0,x^2) + \sum_{n=1}^{\infty} \rho^n B_n(x^0,x^2)\sin(n\phi) + \sum_{n=1}^{\infty} \rho^n (-A_n(x^0,x^2))\cos(n\phi).$$

where $q_1^0$ and $q_3^0$ are components of $|Q^0\rangle$. Similar expressions are for $q_2^0$ and $q_4^0$. Finally,

$$q_1 = q_1^0 - f_1, \quad q_2 = q_2^0 - f_2$$
$$q_3 = q_3^0 - g_1, \quad q_4 = q_4^0 - g_2$$

where $q_i$ are components of $|q\rangle$ and $f_1, f_2, g_1, g_2$ are components of $F$ and $G$ respectively. The coefficients for $q_i^0$ are determined the same way as in the Appendix A, assuming that values of $q_i$ at $\rho = 0$ and $\rho = \rho_0$ are known, etc.

**Acknowledgments**
The results presented in this paper are the outcome of independent research not supported by any institution or government grant.